\pdfoutput=1

\documentclass[11pt]{article}

\usepackage[]{acl}

\usepackage{times}
\usepackage{latexsym}

\usepackage[T1]{fontenc}

\usepackage[utf8]{inputenc}

\usepackage{microtype}

\usepackage{graphicx}
\usepackage{booktabs}
\usepackage{multirow}
\usepackage{multicol}
\usepackage{amsmath}
\usepackage{adjustbox}
\usepackage{subfigure}
\usepackage{hyperref}
\newcommand{\tabitem}{~~\llap{\textbullet}~~}

\title{Multimodal Entity Tagging with Multimodal Knowledge Base}

\author{Hao Peng$^{1}$\thanks{\quad The work was done when Peng was an intern at ByteDance AI Lab.}, Hang Li$^{2}$, Lei Hou$^{1}$, Juanzi Li$^{1}$, Chao Qiao$^{2}$ \\ 
$^1$Department of Computer Science and Technology, Tsinghua University, Beijing, China\\
$^2$ByteDance AI Lab, Beijing, China\\
{\tt peng-h21@mails.tsinghua.edu.cn}
}
\begin{document}
\maketitle

\begin{abstract}

To enhance research on multimodal knowledge base and multimodal information processing, we propose a new task called multimodal entity tagging (MET) with a multimodal knowledge base (MKB). We also develop a dataset for the problem using an existing MKB. In an MKB, there are entities and their associated texts and images. In MET, given a text-image pair, one uses the information in the MKB to automatically identify the related entity in the text-image pair. We solve the task by using the
information retrieval paradigm and implement several baselines using state-of-the-art methods in NLP and CV.  We conduct extensive experiments and make analyses on the experimental results. The results show that the task is challenging, but current technologies can achieve relatively high performance. The data and code are released at https://github.com/h-peng17/MMET.

\end{abstract}

\section{Introduction}

Multimodal knowledge base (MKB) or multimodal knowledge graph (MKG) is an important area for AI technologies because humans' information processing is inherently multimodal. For example, it is believed that humans learn and utilize concepts such as ``Eiffel Tower'' through the processing of multimodal data~\citep{bergen2012louder}. 
Construction and utilization of MKB both need to be intensively investigated.  In this paper, we propose a new task named \textbf{M}ultimodal \textbf{E}ntity \textbf{T}agging (MET) with an MKB and study the problem empirically.

\begin{figure*}
    \centering
    \includegraphics[width=0.98\linewidth]{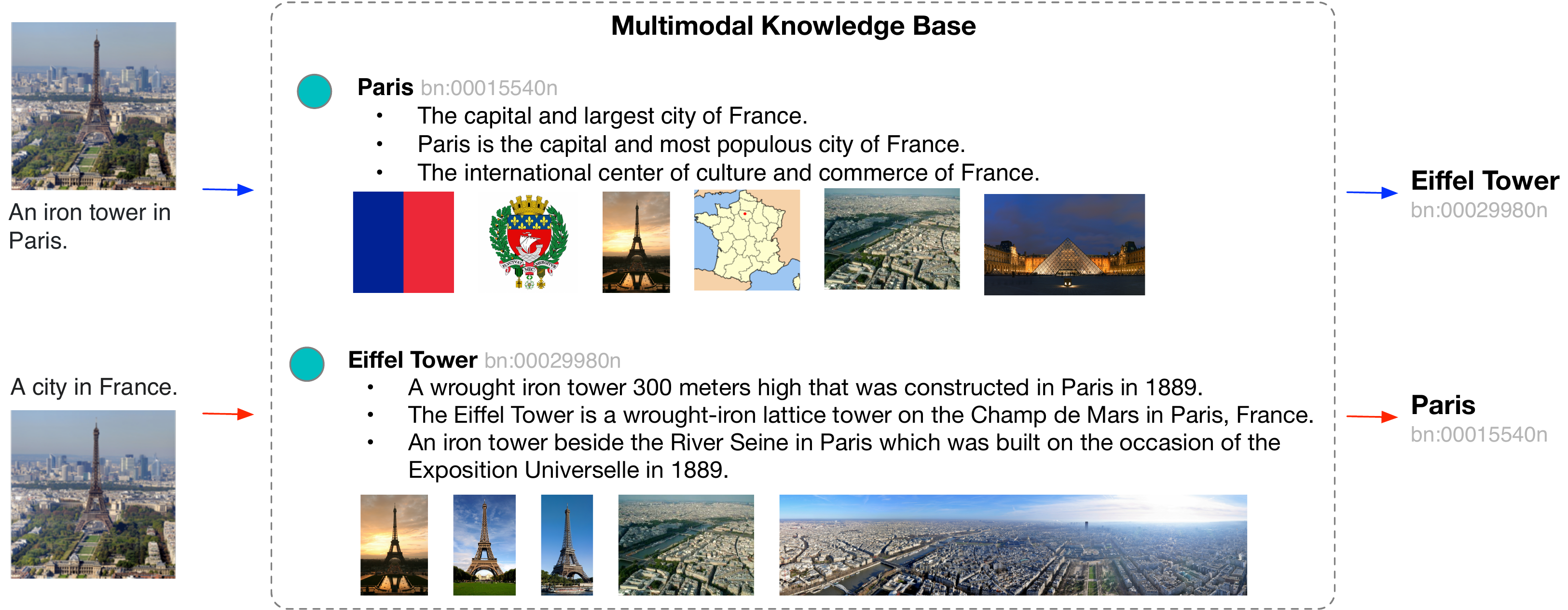}
    \caption{Multimodal entity tagging with multimodal knowledge base. An input contains an image and its text description, such as "An iron tower in Paris". An MKB contains a vast number of entities which have various images and texts associated. We decide whether the input text and image pair describes the entity ``Eiffel Tower'', by taking it as query, and retrieve relevant entities in the MKB, and ranking the candidate entities. 
    }
    \label{fig:archi}
\end{figure*}

Suppose that we have an MKB containing a vast number of entities and each entity has a large number of texts and images associated. Given a new pair of text and image, MET is to identify the entity described in the given text-image pair with an MKB as shown in Figure~\ref{fig:archi}. The task is crucial, we believe, as a step in multimodal information processing.  Note that recognizing whether an image consists of an entity is still a challenging problem in CV~\citep{joseph2021towards}, and here we expect that in addition to an image, its paired text is also used. This is the first work on the issue, as far as we know.

MET is very challenging because of the diversity, sparsity and ambiguity in MKB: 1) Texts and images are diverse. 2) Many entities contain limited information. 3) An image (or a text) may associate with multiple entities. We perform the task by using the information retrieval paradigm. 
Given a pair of text and image, candidates (entities in MKB) are first retrieved using retrieval models. Then we utilize intra-modality and inter-modality matching models to rank the candidates.  We create a dataset of MET from a large MKB called VisualSem~\citep{alberts2020visualsem} and employ state-of-the-art technologies in NLP and CV to conduct extensive experiments on the dataset.
We demonstrate to what extent the existing methods work and provide a foundation for future research on the problem.

%
%
%
%

\section{Task and Dataset}


\subsection{Task Definition}
The task is to identify the entity described in a given text-image pair with an MKB, where entities are depicted in texts and images in the MKB.  
Formally, suppose that there is a multimodal knowledge base having $k$ entities $\mathcal{K} = \{e_1, e_2, \cdots, e_k \}$. Each entity $e_i$ is associated with $i_n$ texts and $i_m$ images, $\mathcal{E}_i = 
\{
\{
t_{1}^{(i)}, t_{2}^{(i)}, 
\cdots,
t_{i_n}^{(i)}
\},
\{
v^{(i)}_{1}, v^{(i)}_{2}, 
\cdots,
v^{(i)}_{i_m}
\}
\}$ where $t_{*}^{(i)}$ and $v^{(i)}_{*}$ denote a text and an image respectively. Further suppose that there is a pair of text and image $(t,v)$. MET aims to find the corresponding entity $e_i \in \mathcal{K}$ that is described by $(t,v)$. 
Here, we expect that a model is automatically learned from the MKB, and the task is performed with the model. 

\begin{table}[h]
    \small
    \centering
    \begin{tabular}{l|rrr}
    \toprule
    \textbf{Data} & \textbf{\# of entities} & \textbf{\# of glosses} & \textbf{\# of images} \\
    \midrule
    MKB & 46,081 & 146,681 & 1,473,574 \\ 
    Train & 46,081 & 46,081 & 46,081  \\ 
    Dev & 1,753 & 1,753 & 1,753 \\
    Test & 1,769 & 1,769 & 1,769 \\
    \bottomrule
    \end{tabular}
    \caption{Statistics of the dataset.}
    \label{tab:statistics}
\end{table}

There are several challenges for the task regarding learning and utilization of the model. First, the scale of the MKB is large, and the content of the MKB is diverse. As shown in Figure~\ref{fig:archi}, images associated with entity ``Paris'' can be landmark buildings, city flag, city emblem, and map. Diversity of the data poses challenges to multimodal information understanding and utilization.
Second, the information might be insufficient for identifying the entities.  For example, many entities in the MKB contain a few images and texts associated.
Third, the data might also be ambiguous. For example, in Figure~\ref{fig:archi}, 
the same input image contains multiple entities (e.g., ``Eiffel Tower'' and ``Paris'') which may confuse the model. Section~\ref{sec:challenge} in appendix provides more details.

\subsection{Dataset Creation}

We derive a new dataset from the multimodal knowledge base VisualSem~\citep{alberts2020visualsem}. VisualSem contains 101,244 entities, and each entity has on average 15.2 images and 2.9 glosses (texts) associated. There is no need to annotate data manually as MKB is \textbf{naturally ``labeled'' data}.
We filter out entities that have less than three glosses or images. We split the data into knowledge-base, training, development, and test sets. We ensure that the entities in the knowledge-base, training, development, and test sets do not have common glosses or images, and thus there is no ``information leak''. Each instance in the training, development, and test sets consists of a randomly combined pair of gloss and image of an entity. Table~\ref{tab:statistics} shows the statistics of the data.



\section{Method}
We view MET with MKB as an information retrieval problem. The input is a text-image pair $(t, v)$ as the query. 1) \textbf{Retrieval} We first separately retrieve the most relevant $N$ texts and their entities with text $t$, and the most relevant $M$ images and their entities with image $v$. Thus, there are at most $N+M$ entities. Next, each retrieved text/image is paired with some image/text of the same entity.\footnote{We give details of this procedure in section~\ref{sec:retrieval} in appendix.\label{fn:repeat}} 2) \textbf{Ranking} We then rank the entities based on the relevance between the query $(t, v)$ and the retrieved texts and images. There are multiple matching scores between the query and the retrieved texts and images, and all of them are taken as features of the ranking model. Finally, the top-ranked entities are selected as output.

\subsection{Retrieval}

\paragraph{Text Retrieval}

We employ the elastic search system \footnote{https://www.elastic.co/} to conduct term-based text retrieval with the text $t$ as query. Once the most relevant $N$ texts 
$
\{
t_{j_1}^{(i_1)},
\cdots,
t_{j_N}^{(i_N)}
\}
$
are retrieved, the corresponding $N$ entities
$
\{
e_{i_1},
\cdots,
e_{i_N}
\}
$
are obtained and $N$ images 
$
\{
v_{\hat{j}_1}^{(i_1)},
\cdots,
v_{\hat{j}_N}^{(i_N)}
\}
$ are selected to pair with retrieved texts.\footnotemark[2]

\paragraph{Image Retrieval}
We utilize a ResNet152 model~\citep{he2016deep} to encode all images as real-valued vectors of dimension 2048. We employ the nearest neighbor search technique\footnote{https://github.com/nmslib/hnswlib} to perform vector-based image retrieval with the image $v$ as query. Once the most relevant $M$ images 
$
\{
v_{\hat{k}_1}^{(l_1)}, 
\cdots,
v_{\hat{k}_M}^{(l_M)}
\}
$
are retrieved, the corresponding $M$ entities
$
\{
e_{l_1}, 
\cdots,
e_{l_M}
\}
$
are obtained and $M$ texts 
$
\{
t_{k_1}^{(l_1)}, 
\cdots,
t_{k_M}^{(l_M)}
\}
$ are selected to pair with retrieved images.\footnotemark[2]

\subsection{Ranking}

The most relevant entities are ranked at the top based on the relevance between the query and the retrieved texts and images.  The ranking model is a linear combination of matching scores where the weights are tuned using the development data. The matching scores are calculated with different matching models. Formally, $\mathbf{e_o} = F((t,v), (t_{p}^{o}, v_{\hat{p}}^{o}))$, $\mathbf{e_o}$ denotes the matching score and $F$ denotes the matching model.

\paragraph{Intra-modality Matching}
1) Text Bi-encoder Matching (\textbf{TBM}).
The bi-encoder model transforms the query text and a retrieved text into their representations with two encoders and calculates the similarity (relevance) between the representations~\citep{wu2020scalable, thakur2021augmented}. The bi-encoder model is usually more efficient. We train the two encoders by fine-tuning a BERT model~\citep{devlin2019bert}. The two encoders share tied parameters as in~\citet{reimers2019sentence}. 2) Text Cross-encoder Matching (\textbf{TCM}). 
The cross-encoder transforms the concatenation of the query text and a retrieved text into a representation with only one encoder~\citep{wu2020scalable}, and decides the matching degree (relevance) between the two texts. The cross-encoder model is more accurate, because interactions between the two texts are captured in the encoder. We train the encoder by fine-tuning the BERT model. 3) Image Bi-encoder Matching (\textbf{IBM}).
The ``image bi-encoder matching'' model has two encoders (tied parameters), one for encoding the query image and the other for encoding a retrieved image. It uses cosine to represent the similarity (relevance) between the representations from the two encoders. We implement each of the two encoders using ResNet152~\citep{he2016deep} pre-trained on ImageNet~\citep{deng2009imagenet}. 

\paragraph{Inter-modality Matching} \textbf{CLIP} is an 
image-text matching model proposed by \citet{DBLP:conf/icml/RadfordKHRGASAM21}.
CLIP is pre-trained on 400 million image text pairs and demonstrates strong performances on downstream image classification tasks, especially in few-shot or zero-shot settings. We adopt CLIP\footnote{https://github.com/openai/CLIP} as a method for inter-modality matching. 

%

\section{Experiments}

\begin{table}[]
    \small
    \centering
    
    \begin{adjustbox}{width=0.99\linewidth}
    
    \begin{tabular}{l|l|rrr}
    \toprule
    \textbf{Stage} & \textbf{Model} & \textbf{Hits@1} & \textbf{Hits@3} & \textbf{Hits@10} \\
    \midrule
    \multirow{2}{*}{Retrieval} & 
    Text  & 41.4 & 51.3 & 62.5 \\
    & Image &7.8 & 11.8 & 17.1 \\ 
    \midrule
    \multirow{5}{*}{Ranking} & 
    TBM & 41.5& 52.9&66.8 \\
    & TCM & 58.4& 69.4 & 78.0 \\
    & IBM &9.3 &12.5 &17.1 \\ 
    & CLIP &16.3 &27.9 &45.3 \\ 
    \cmidrule{2-5} 
    & Full Model & \textbf{61.2} & \textbf{71.4}& \textbf{79.4} \\
     \bottomrule
    \end{tabular}
    \end{adjustbox}
    
    \caption{Experiment results (\%). Full Model utilizes all matching (TBM, TCM, IBM, CLIP) scores to rank entities.}
    \label{tab:experiment}
\end{table}

\subsection{Experimental Settings and Results}

We conduct experiments to investigate the hardness of the MET problem as well as the performance of the methods described above. We use Hits@N as evaluation measure, which is the percentage of correct entities at the top N positions. In our experiments, we set $M = N = 100$. We use grid search to tune the weights of ranking model (Full Model) with the development set.

Table~\ref{tab:experiment} shows the experimental results. The Hits@1 score of image retrieval is only of 7.8\%, indicating that using image data alone would not achieve high performance in retrieval. This is due to the diversity and ambiguity of images. In contrast, the Hits@1 of text retrieval is as high as 41.4\%, which indicates that it is more effective to use text data to carry out retrieval.

Table~\ref{tab:experiment} also shows the results of ranking. We make the following observations. 
1) The results indicate that the full model of using TCM, TBM, IBM, and CLIP as matching models performs the best in terms of Hits@1. 2) The text cross-encoder matching model (TCM) makes a large performance improvement after the retrieval. The result indicates that the texts in the MKB contain more information and the use of text data is essential for MET. 
3) The image bi-encoder matching model (IBM) makes a small improvement after the retrieval, because of the diversity and ambiguity of images. It appears that the training of IBM is challenging, and the model is confused by the training data. 4) CLIP achieves a relatively low performance. Although CLIP works remarkably well in zero-shot image classification~\cite{DBLP:conf/icml/RadfordKHRGASAM21}, it still under-performs a text matching method. The result indicates that we still need to enhance the capability of the CLIP model.

\begin{table}
    \small
    \centering
    \begin{tabular}{l|rrr}
    \toprule
    \textbf{Model} & \textbf{Hit@1} & \textbf{Hits@3} & \textbf{Hits@10} \\ 
    \midrule
    Full Model & \textbf{61.2} & \textbf{71.4} & \textbf{79.4} \\ 
    \midrule
    w/o IBM & 59.8 & 70.6 & 79.0 \\ 
    w/o CLIP & 60.0 & 70.5 & 78.9 \\
    \midrule
    w/o TBM & 60.0 &70.8  &79.1 \\
    w/o TCM &43.5 & 57.1 & 69.8 \\ 
    \bottomrule
    \end{tabular}
    \caption{Ablation study results (\%).}
    \label{tab:ablation}
\end{table}

\subsection{Ablation Study}
We conduct an ablation study and examine the contributions from different matching models. As shown in Table~\ref{tab:ablation}, all models make contributions and the performance will drop if any of them is removed. Text information is essential, as excluding it (\textbf{w/o TBM} or \textbf{w/o TCM}) brings a significant performance decrease. Though identifying entities in images is challenging (9.3\% Hits@1 for IBM), images still provide helpful information in multimodal entity tagging, because excluding image information (\textbf{w/o IBM} or \textbf{w/o CLIP}) hurts the performance. In conclusion, the task needs multimodal information and powerful multimodal models.

\subsection{Error Analysis}
We randomly sample 100 text-image pairs that are incorrectly tagged. We find three types of errors: ``Noisy'', ``Hard'' and ``Wrong''. ``Noisy'' means that
noise in the dataset misleads models. ``Hard'' means that it is not easy for the models to resolve the ambiguity of texts (e.g., texts are general and simple) and images (e.g., entities in images are rare). ``Wrong'' means that the multimodal information in the input pair and MKB is clear but models fail to utilize the information to recognize entities. It turns out that 46\% of the errors are Hard cases, 42\% are Wrong cases, and 12\% are Noisy cases. This indicates that the state-of-the-art models still cannot accomplish the task satisfactorily.

\section{Related Work}

There are several multimodal knowledge bases.
\citet{DBLP:conf/ijcai/XieLLS17} create WN9-IMG, which consists of a subset entities of WordNet~\citep{miller1995wordnet} and images from ImageNet~\citep{deng2009imagenet}. \citet{mousselly2018multimodal} develop FB-IMG, which consists of entities from Freebase~\citep{bollacker2008freebase} and images from the web. \citet{liu2019mmkg} construct MMKG, containing three knowledge bases DBpedia15K, YAGO15K and Freebase15K. 
\citet{wang2020richpedia} build Richpedia, a large-scale 
multimodal knowledge base, which consists of ``KG entities'' and ``image entities'' on three topics. 
Recently, \citet{alberts2020visualsem} develop a multimodal and multilingual knowledge base named VisualSem. VisualSem consists of 101,244 entities, 1,539,244 images and multilingual texts. We derive the new dataset from VisualSem because it is the largest MKB publically available. 

There is also existing work on multimodal entity linking~\citep{moon2018multimodal, adjali2020building, adjali2020multimodal,  zhang2021attention}, which manages to link entities mentioned in texts using image data. For example,
\citet{moon2018multimodal} introduce multimodal named entity disambiguation (MNED), which leverages visual contexts for entity linking in texts in social media.
\citet{adjali2020building,adjali2020multimodal} publish a multimodal entity linking dataset and utilize a combination of text, BM25, popularity, visual features to link entities in tweet data. \citet{zhang2021attention} propose a new attention-based multimodal entity linking method and construct a new Chinese multimodal entity linking data set based on Weibo (https://weibo.com/).




\section{Conclusion}
We propose multimodal entity tagging (MET) with multimodal knowledge base (MKB), which is to identify the most related entity in a given text and image pair, using the information in an MKB. The new task is important for enhancing research on construction and utilization of MKB. We solve the problem by using the information retrieval paradigm. We construct a new large-scale dataset for the task and conduct intensive experiments.
Experimental results indicate that the task is still challenging, and more powerful models for multimodal representation learning are needed. 

\bibliography{anthology,custom}

\begin{thebibliography}{22}
\expandafter\ifx\csname natexlab\endcsname\relax\def\natexlab#1{#1}\fi

\bibitem[{Adjali et~al.(2020{\natexlab{a}})Adjali, Besan{\c{c}}on, Ferret,
  Le~Borgne, and Grau}]{adjali2020building}
Omar Adjali, Romaric Besan{\c{c}}on, Olivier Ferret, Herv{\'e} Le~Borgne, and
  Brigitte Grau. 2020{\natexlab{a}}.
\newblock Building a multimodal entity linking dataset from tweets.
\newblock In \emph{Proceedings of the 12th Language Resources and Evaluation
  Conference}, pages 4285--4292.

\bibitem[{Adjali et~al.(2020{\natexlab{b}})Adjali, Besan{\c{c}}on, Ferret,
  Le~Borgne, and Grau}]{adjali2020multimodal}
Omar Adjali, Romaric Besan{\c{c}}on, Olivier Ferret, Herv{\'e} Le~Borgne, and
  Brigitte Grau. 2020{\natexlab{b}}.
\newblock Multimodal entity linking for tweets.
\newblock \emph{Advances in Information Retrieval}, 12035:463.

\bibitem[{Alberts et~al.(2020)Alberts, Huang, Deshpande, Liu, Cho, Vania, and
  Calixto}]{alberts2020visualsem}
Houda Alberts, Teresa Huang, Yash Deshpande, Yibo Liu, Kyunghyun Cho, Clara
  Vania, and Iacer Calixto. 2020.
\newblock Visualsem: a high-quality knowledge graph for vision and language.
\newblock \emph{arXiv preprint arXiv:2008.09150}.

\bibitem[{Bergen(2012)}]{bergen2012louder}
Benjamin~K Bergen. 2012.
\newblock \emph{Louder than words: The new science of how the mind makes
  meaning}.
\newblock Basic Books (AZ).

\bibitem[{Bollacker et~al.(2008)Bollacker, Evans, Paritosh, Sturge, and
  Taylor}]{bollacker2008freebase}
Kurt Bollacker, Colin Evans, Praveen Paritosh, Tim Sturge, and Jamie Taylor.
  2008.
\newblock Freebase: a collaboratively created graph database for structuring
  human knowledge.
\newblock In \emph{Proceedings of the 2008 ACM SIGMOD international conference
  on Management of data}, pages 1247--1250.

\bibitem[{Deng et~al.(2009)Deng, Dong, Socher, Li, Li, and
  Fei-Fei}]{deng2009imagenet}
Jia Deng, Wei Dong, Richard Socher, Li-Jia Li, Kai Li, and Li~Fei-Fei. 2009.
\newblock Imagenet: A large-scale hierarchical image database.
\newblock In \emph{2009 IEEE conference on computer vision and pattern
  recognition}, pages 248--255. Ieee.

\bibitem[{Devlin et~al.(2019)Devlin, Chang, Lee, and
  Toutanova}]{devlin2019bert}
Jacob Devlin, Ming-Wei Chang, Kenton Lee, and Kristina Toutanova. 2019.
\newblock Bert: Pre-training of deep bidirectional transformers for language
  understanding.
\newblock In \emph{Proceedings of the 2019 Conference of the North American
  Chapter of the Association for Computational Linguistics: Human Language
  Technologies, Volume 1 (Long and Short Papers)}, pages 4171--4186.

\bibitem[{He et~al.(2016)He, Zhang, Ren, and Sun}]{he2016deep}
Kaiming He, Xiangyu Zhang, Shaoqing Ren, and Jian Sun. 2016.
\newblock Deep residual learning for image recognition.
\newblock In \emph{Proceedings of the IEEE conference on computer vision and
  pattern recognition}, pages 770--778.

\bibitem[{Joseph et~al.(2021)Joseph, Khan, Khan, and
  Balasubramanian}]{joseph2021towards}
KJ~Joseph, Salman Khan, Fahad~Shahbaz Khan, and Vineeth~N Balasubramanian.
  2021.
\newblock Towards open world object detection.
\newblock In \emph{Proceedings of the IEEE/CVF Conference on Computer Vision
  and Pattern Recognition}, pages 5830--5840.

\bibitem[{Lin et~al.(2017)Lin, Goyal, Girshick, He, and
  Doll{\'a}r}]{lin2017focal}
Tsung-Yi Lin, Priya Goyal, Ross Girshick, Kaiming He, and Piotr Doll{\'a}r.
  2017.
\newblock Focal loss for dense object detection.
\newblock In \emph{Proceedings of the IEEE international conference on computer
  vision}, pages 2980--2988.

\bibitem[{Liu et~al.(2019)Liu, Li, Garcia-Duran, Niepert, Onoro-Rubio, and
  Rosenblum}]{liu2019mmkg}
Ye~Liu, Hui Li, Alberto Garcia-Duran, Mathias Niepert, Daniel Onoro-Rubio, and
  David~S Rosenblum. 2019.
\newblock Mmkg: multi-modal knowledge graphs.
\newblock In \emph{European Semantic Web Conference}, pages 459--474. Springer.

\bibitem[{Miller(1995)}]{miller1995wordnet}
George~A Miller. 1995.
\newblock Wordnet: a lexical database for english.
\newblock \emph{Communications of the ACM}, 38(11):39--41.

\bibitem[{Moon et~al.(2018)Moon, Neves, and Carvalho}]{moon2018multimodal}
Seungwhan Moon, Leonardo Neves, and Vitor Carvalho. 2018.
\newblock Multimodal named entity disambiguation for noisy social media posts.
\newblock In \emph{Proceedings of the 56th Annual Meeting of the Association
  for Computational Linguistics (Volume 1: Long Papers)}, pages 2000--2008.

\bibitem[{Mousselly-Sergieh et~al.(2018)Mousselly-Sergieh, Botschen, Gurevych,
  and Roth}]{mousselly2018multimodal}
Hatem Mousselly-Sergieh, Teresa Botschen, Iryna Gurevych, and Stefan Roth.
  2018.
\newblock A multimodal translation-based approach for knowledge graph
  representation learning.
\newblock In \emph{Proceedings of the Seventh Joint Conference on Lexical and
  Computational Semantics}, pages 225--234.

\bibitem[{Radford et~al.(2021)Radford, Kim, Hallacy, Ramesh, Goh, Agarwal,
  Sastry, Askell, Mishkin, Clark, Krueger, and
  Sutskever}]{DBLP:conf/icml/RadfordKHRGASAM21}
Alec Radford, Jong~Wook Kim, Chris Hallacy, Aditya Ramesh, Gabriel Goh,
  Sandhini Agarwal, Girish Sastry, Amanda Askell, Pamela Mishkin, Jack Clark,
  Gretchen Krueger, and Ilya Sutskever. 2021.
\newblock \href {http://proceedings.mlr.press/v139/radford21a.html} {Learning
  transferable visual models from natural language supervision}.
\newblock In \emph{Proceedings of the 38th International Conference on Machine
  Learning, {ICML} 2021, 18-24 July 2021, Virtual Event}, volume 139 of
  \emph{Proceedings of Machine Learning Research}, pages 8748--8763. {PMLR}.

\bibitem[{Reimers and Gurevych(2019)}]{reimers2019sentence}
Nils Reimers and Iryna Gurevych. 2019.
\newblock Sentence-bert: Sentence embeddings using siamese bert-networks.
\newblock In \emph{Proceedings of the 2019 Conference on Empirical Methods in
  Natural Language Processing and the 9th International Joint Conference on
  Natural Language Processing (EMNLP-IJCNLP)}, pages 3982--3992.

\bibitem[{Thakur et~al.(2021)Thakur, Reimers, Daxenberger, and
  Gurevych}]{thakur2021augmented}
Nandan Thakur, Nils Reimers, Johannes Daxenberger, and Iryna Gurevych. 2021.
\newblock Augmented sbert: Data augmentation method for improving bi-encoders
  for pairwise sentence scoring tasks.
\newblock In \emph{Proceedings of the 2021 Conference of the North American
  Chapter of the Association for Computational Linguistics: Human Language
  Technologies}, pages 296--310.

\bibitem[{Wang et~al.(2020)Wang, Wang, Qi, and Zheng}]{wang2020richpedia}
Meng Wang, Haofen Wang, Guilin Qi, and Qiushuo Zheng. 2020.
\newblock Richpedia: a large-scale, comprehensive multi-modal knowledge graph.
\newblock \emph{Big Data Research}, 22:100159.

\bibitem[{Wolf et~al.(2020)Wolf, Debut, Sanh, Chaumond, Delangue, Moi, Cistac,
  Rault, Louf, Funtowicz, Davison, Shleifer, von Platen, Ma, Jernite, Plu, Xu,
  Scao, Gugger, Drame, Lhoest, and Rush}]{wolf-etal-2020-transformers}
Thomas Wolf, Lysandre Debut, Victor Sanh, Julien Chaumond, Clement Delangue,
  Anthony Moi, Pierric Cistac, Tim Rault, Rémi Louf, Morgan Funtowicz, Joe
  Davison, Sam Shleifer, Patrick von Platen, Clara Ma, Yacine Jernite, Julien
  Plu, Canwen Xu, Teven~Le Scao, Sylvain Gugger, Mariama Drame, Quentin Lhoest,
  and Alexander~M. Rush. 2020.
\newblock \href {https://www.aclweb.org/anthology/2020.emnlp-demos.6}
  {Transformers: State-of-the-art natural language processing}.
\newblock In \emph{Proceedings of the 2020 Conference on Empirical Methods in
  Natural Language Processing: System Demonstrations}, pages 38--45, Online.
  Association for Computational Linguistics.

\bibitem[{Wu et~al.(2020)Wu, Petroni, Josifoski, Riedel, and
  Zettlemoyer}]{wu2020scalable}
Ledell Wu, Fabio Petroni, Martin Josifoski, Sebastian Riedel, and Luke
  Zettlemoyer. 2020.
\newblock Scalable zero-shot entity linking with dense entity retrieval.
\newblock In \emph{Proceedings of the 2020 Conference on Empirical Methods in
  Natural Language Processing (EMNLP)}, pages 6397--6407.

\bibitem[{Xie et~al.(2017)Xie, Liu, Luan, and Sun}]{DBLP:conf/ijcai/XieLLS17}
Ruobing Xie, Zhiyuan Liu, Huanbo Luan, and Maosong Sun. 2017.
\newblock \href {https://doi.org/10.24963/ijcai.2017/438} {Image-embodied
  knowledge representation learning}.
\newblock In \emph{Proceedings of the Twenty-Sixth International Joint
  Conference on Artificial Intelligence, {IJCAI} 2017, Melbourne, Australia,
  August 19-25, 2017}, pages 3140--3146. ijcai.org.

\bibitem[{Zhang et~al.(2021)Zhang, Li, and Yang}]{zhang2021attention}
Li~Zhang, Zhixu Li, and Qiang Yang. 2021.
\newblock Attention-based multimodal entity linking with high-quality images.
\newblock In \emph{International Conference on Database Systems for Advanced
  Applications}, pages 533--548. Springer.

\end{thebibliography}
\bibliographystyle{acl_natbib}

\appendix
\section{Training Details}
We implement the text matching methods using HuggingFace transformers library~\citep{wolf-etal-2020-transformers} and the image matching methods using torchvision\footnote{https://github.com/pytorch/vision}. We utilize BERT-Base (\texttt{bert-base-uncased}) as text encoder and ResNet152 as image encoder. For CLIP, we use ViT-B/16 as the backbone. 
We take the new task as an information retrieval problem. 
First, two retrievers separately retrieve 100 entities.\footnote{The parameters of retrievers are not tuned on the training data.} Due to the Cuda memory limits, we next sample 64 out of 200 candidates. The batch size is one that contains only one input text-image pair and 64 retrieved candidates. The number of positive and negative candidates is unbalanced, and thus we ensure that at least one positive candidate is sampled in each batch and utilize the focal loss~\citep{lin2017focal} as training objective. 
We utilize AdamW to optimize the text encoder and SGD to optimize the image encoder for efficient training. The learning rates are 3e-5 and 1e-2, respectively. All models are trained in 20 epochs on eight NVIDIA Tesla V100 GPUs. 
We conduct grid search (in $\{0, 0.1, \cdots, 0.9\}$) to tune the weights of ranking model (Full Model) with the development set. The weights of TCM, TBM, IBM, CLIP matching scores are $0.2$, $0.1$, $0.3$, $0.9$ respectively.

\section{Retrieval Details \label{sec:retrieval}}
We use text retrieval model to retrieve $N$ entities 
$
\{
e_{i_1}, e_{i_2}, 
\cdots,
e_{i_N}
\}
$ with $
\{
t_{j_1}^{(i_1)}, t_{j_2}^{(i_2)}, 
\cdots,
t_{j_N}^{(i_N)}
\}
$ associated. And we use image retrieval model to 
retrieve $M$ entities 
$
\{
e_{l_1}, e_{l_2}, 
\cdots,
e_{l_M}
\}
$ with $
\{
v_{\hat{k}_1}^{(l_1)}, v_{\hat{k}_2}^{(l_2)}, 
\cdots,
v_{\hat{k}_M}^{(l_M)}
\}
$ associated. After retrieval, $N + M$ entities 
$
\{
e_{i_1},
\cdots,
e_{i_N},
e_{l_1},
\cdots,
e_{l_M}
\}
$ are retrieved. However, images are not retrieved in text retrieval procedure and vice versa. We select an image $v_{\hat{j}}^{(i)}$ of the same entity for each $t_{j}^{(i)}$ in $N$ retrieved texts randomly at training time and select an image $v_{0}^{(i)}$ (i.e., the first image of the entity)\footnote{We find the first image (or text)  of an entity contains little noise.} at inference time. The same procedure is used to select texts for retrieved images. The final retrieval results are $
\{
e_{i_1},
\cdots,
e_{i_N},
e_{l_1},
\cdots,
e_{l_M}
\}
$ with 
$
\{
(t_{j_1}^{(i_1)}, v_{\hat{j}_1}^{(i_1)}), 
\cdots,
(t_{j_N}^{(i_N)}, v_{\hat{j}_N}^{(i_N)}), 
\cdots,
(t_{k_M}^{(l_M)}, v_{\hat{k}_M}^{(l_M)})
\}
$ associated.

\section{Challenges \label{sec:challenge}}
We show the challenges of MET in diversity, sparsity, ambiguity of the data in MKB.
\subsection{Diversity}
Entities usually have various texts and images associated. The diversity of data brings challenges to multimodal information understanding. Table~\ref{tab:diversity} shows three entities with diverse texts and images. 
``Romanesque Architecture'' is a concept and contains various images, including different architectures and graphics. ``Paris'' is an entity that also has various images associated. The images of ``Paris'' can be landmark buildings (e.g., ``Eiffel Tower'',
``Louvre'', ``Arc de Triomphe''), city flag, city emblem, and map. Usually, 
These types of entities (place and organization) have various images associated, which poses challenges. ``Francisco de Goya'' belongs to another type. ``Francisco de Goya'' is a painter and thus has many images of artworks associated.

\subsection{Data Sparsity}
Many entities in the multimodal knowledge base do not have sufficient information. About 9.7\% of the entities contain no more than three images in the dataset and 12.2\% of the entities contain only one text. We conduct experiments on sparse text and sparse image data.
We can observe from the results in Table~\ref{tab:spa_result}. 1) On sparse text data, the performance of text matching models (TBM and TCM) drops significantly. 2) On sparse image data, the performance of image matching model (IBM) also drops significantly. Hits@10 (6.5\%) is even lower than Hits@1 (9.3\%).
3) The performance of the inter-modality matching model CLIP deteriorates on sparse data as well, especially on sparse image data. In conclusion, sparsity is a challenge in the dataset and more data-efficient models are needed.

\subsection{Ambiguity}
An image may contain several entities. 
About 21.6\% of the images in the dataset are associated with two or more entities. 
Furthermore, entities in images may be rare. Therefore, recognizing entities from the images using visual information alone is challenging. Table~\ref{tab:ambiguity} shows examples indicating the ambiguity of images. 
The animal in the first image is ``Genet''. However, considering more general
species categories, it can be ``Viverrine'' or ``Chordate'' (``Procyon'' is a mistakenly associated entity). There are several entities in the second and third images. Depending on the texts, 
the recognized entities might be different. The entities in the third image are rare. Therefore, the utilization of texts is essential to 
recognize the entities in query.  In the meantime, texts are also ambiguous. For example, the descriptions are general and simple.  Table~\ref{tab:text_am} gives the top ten
frequent texts, and the texts are ambiguities. Although ambiguous texts are not common (1.7\% of the texts in the MKB have two and more entities associated), it still brings challenges to the task.

\begin{table}
    \small
    \centering
    \begin{tabular}{lr}
        \toprule
        \textbf{Text} & \textbf{\# of Entities} \\ 
        \midrule
        A province of Indonesia & 25\\
        One of the moons of Jupiter & 20 \\
        State of Mexico & 19 \\ 
        Disease & 19 \\
        Genus of reptiles (fossil) & 19 \\ 
        A city of Japan & 14 \\ 
        American musician & 14 \\ 
        Year & 13 \\ 
        ISO 3166-1 country code & 13 \\ 
        Medical specialty & 13 \\ 
        Male given name & 13 \\
        \bottomrule
    \end{tabular}
    \caption{Ambiguity of texts. The texts are general, simple and thus ambiguous.}
    \label{tab:text_am}
\end{table}

\begin{table}[h]
    \small
    \centering
    
    \begin{adjustbox}{width=0.99\linewidth}
    
    \begin{tabular}{l|rrr}
    \toprule
    \textbf{Model} & \textbf{Hits@1} & \textbf{Hits@3} & \textbf{Hits@10} \\
    \midrule
    TBM & 41.5& 52.9&66.8 \\
    \quad w/ assemble &41.0 &49.5 &64.2 \\
    TCM & 58.4& 69.4& 78.0 \\
    \quad w/ assemble & \underline{59.4} & 67.2& \underline{78.3} \\
    IBM &9.3 &12.5 &17.1 \\ 
    \quad w/ assemble & 4.6 & 7.1 & 12.7 \\
    CLIP &16.3 &27.9 &45.3 \\ 
    \quad w/ assemble & \underline{20.6} & \underline{31.0} & \underline{48.2} \\
    \midrule 
    Full Model & 61.2 & \textbf{71.4}& 79.4 \\
    \quad w/ assemble &\underline{\textbf{63.9}} &70.9 & \underline{\textbf{79.5}} \\
     \bottomrule
    \end{tabular}
    
    \end{adjustbox}
    
    \caption{Assembling results (\%). \underline{Underline}s indicate assembling improves the performance. }
    \label{tab:assemble}
\end{table}

\section{Experimental Results}
\subsection{Assembling of Ranking Model}
After retrieval, each retrieved text/image is paired with the first image/text of the same entity at inference time as mentioned in Section~\ref{sec:retrieval}. However, considering the diversity of MKB, assembling more instances (an instance is a text or an image) when computing matching scores for entities is a straightforward idea to enhance the performance. We conduct additional experiments to assemble multiple instances in \textbf{ranking} procedure at inference time. Specifically, for each candidate $e_o$ in retrieved entities
$\{
e_{i_1},
\cdots,
e_{i_N},
e_{l_1},
\cdots,
e_{l_M}
\}
$, we select $K$ instances including 
the retrieved instance ($v_{\hat{p}}^{(o)}$ or $t_{p}^{(o)}$) to compute matching scores with input instance (image $v$ or text $t$ considering different matching models) and then average them as the final score. The assemble of Full Model is linear combination of separate assemble models and the weights of TCM, TBM, IBM, CLIP matching scores are $0.1$, $0.3$, $0.1$, $0.9$ respectively.\footnote{We conduct grid search with development set.}
We set $K=3$ in our experiments and larger $K$ may bring further improvement which we leave as future work.

Table~\ref{tab:assemble} illustrates the assembling results. We can observe that: 1) Hits@1 of assembled ranking model (Full Model) improves significantly, indicating more images (or texts) bring more information to distinguish and recognize the corresponding entities. 
2) Hits@1 of CLIP improves a lot, indicating inter-modality matching models benefit more from the diversity of MKB.  3) The assemble performance of IBM and TBM drops and Hits@1 of IBM drops about 50\% compared to original Hits@1.
The main reason is that the additional instances (images or texts) \textbf{may not be retrieved by retrieval models} which means the relevance between instances and entities can not be learned easily,  especially for bi-encoder models which are less powerful and more difficult to train. 

In conclusion, utilization of more instances (images or texts) may bring performance improvement to models especially inter-modality matching models due to the diversity of MKB. Meanwhile, 
more instances may confuse models and hurt the performance, especially for less powerful bi-encoder matching models.

\subsection{More Experimental Results}
Table~\ref{tab:overall} reports the results on both development and test sets. One can see that the results on development and test sets are generally consistent.  Hits@100 achieves 86\%, about 20\% higher than Hits@1, indicating that there
is still room for performance improvement and more powerful models for multimodal representation are needed.


\begin{table*}
    \small
    \centering
    \begin{tabular}{l|rrr|rrr}
        \toprule
        \multirow{2}{*}{\textbf{Model}} & 
        \multicolumn{3}{c|}{\textbf{Sparse Text Data}} & 
        \multicolumn{3}{c}{\textbf{Sparse Image Data}} \\ 
        & \textbf{Hits@1} & \textbf{Hits@3} & \textbf{Hits@10} 
        & \textbf{Hits@1} & \textbf{Hits@3} & \textbf{Hits@10} \\
        \midrule 
        TBM & 30.5{\tiny \color{blue} -$11.0$} & 40.3{\tiny \color{blue} -$12.6$}
        & 54.5{\tiny \color{blue} -$12.3$}
        & 41.1{\tiny \color{blue} -$0.4$}
        & 51.2{\tiny \color{blue} -$1.7$}
        & 61.7{\tiny \color{blue} -$5.1$}{\tiny \color{white} $0$} \\
        TCM & 46.2{\tiny \color{blue} -$12.2$} 
        & 57.3{\tiny \color{blue} -$12.1$} 
        & 67.5{\tiny \color{blue} -$10.5$}
        & 55.2{\tiny \color{blue} -$3.2$}
        & 66.1{\tiny \color{blue} -$3.3$} 
        & 74.2{\tiny \color{blue} -$3.8$}{\tiny \color{white} $0$} \\
        IBM & 8.3{\tiny \color{blue} -$1.0$}{\tiny \color{white} $0$}  
        & 12.3{\tiny \color{blue} -$0.2$}{\tiny \color{white} $0$}  
        & 17.2{\tiny \color{red} +$0.1$}{\tiny \color{white} $0$}  
        & 4.4{\tiny \color{blue} -$4.9$} 
        & 5.2{\tiny \color{blue} -$7.3$}
        & 6.5{\tiny \color{blue} -$10.6$}\\
        CLIP & 15.3{\tiny \color{blue} -$1.0$}{\tiny \color{white} $0$}  
        & 25.0{\tiny \color{blue} -$2.9$}{\tiny \color{white} $0$}  
        & 39.7{\tiny \color{blue} -$5.6$}{\tiny \color{white} $0$}  
        & 11.7{\tiny \color{blue} -$4.6$} 
        & 19.8{\tiny \color{blue} -$8.1$} 
        & 32.7{\tiny \color{blue} -$12.6$} \\
        \midrule
        Full Model & 50.3{\tiny \color{blue} -$10.9$} 
        & 60.6{\tiny \color{blue} -$10.8$} 
        & 69.4{\tiny \color{blue} -$10.0$} 
        & 58.1{\tiny \color{blue} -$3.1$} 
        & 66.5{\tiny \color{blue} -$4.9$} 
        & 75.0{\tiny \color{blue} -$4.4$}{\tiny \color{white} $0$}  \\
        \bottomrule
    \end{tabular}
    \caption{Results on sparse data (\%). All models perform worse on sparse data, especially on sparse text data. This shows sparsity is still a challenge to the new task MET. }
    \label{tab:spa_result}
\end{table*}

\begin{table*}[]
    \small
    \centering
    \begin{tabular}{l|l|rrrr|rrrr}
    \toprule
    \multirow{2}{*}{\textbf{Stage}} & 
    \multirow{2}{*}{\textbf{Model}} & 
    \multicolumn{4}{c|}{\textbf{Dev}} & 
    \multicolumn{4}{c}{\textbf{Test}} \\
    & & \textbf{Hits@1} & \textbf{Hits@3} & \textbf{Hits@10} & \textbf{Hits@100} & \textbf{Hits@1} & \textbf{Hits@3} & \textbf{Hits@10} & \textbf{Hits@100} \\
    \midrule
    \multirow{2}{*}{Retrieval} & 
    Text & 38.8 & 50.4 & 63.1 & 79.6 & 41.4 & 51.3 & 62.5 & 78.9 \\
    & Image & 7.9 &12.7  & 19.1 & 32.6 &7.8 & 11.8 & 17.1 & 31.3 \\ 
    \midrule
    \multirow{5}{*}{Ranking} & 
    TBM & 41.0 & 52.3 & 65.0 & 82.1 & 41.5& 52.9&66.8 & 81.7 \\
    & TCM & 59.0 & 70.2 & 78.5 & 85.9 & 58.4& 69.4& 78.0 & 85.4 \\
    & IBM &10.5 &14.0 &18.9 &34.0 &9.3 &12.5 &17.1 & 34.1 \\ 
    & CLIP & 18.5 &29.9 &46.6 &82.9 &16.3 &27.9 &45.3 & 82.3 \\ 
    \cmidrule{2-10} 
    & Full Model & \textbf{61.6} & \textbf{71.7} & \textbf{79.6} & 
    \textbf{86.1} & \textbf{61.2} & \textbf{71.4}& \textbf{79.4} & \textbf{85.5}\\
     \bottomrule
    \end{tabular}
    \caption{All experiment results (\%). The results on development and test sets are generally consistent which indicates that development set is suitable for model selection. }
    \label{tab:overall}
\end{table*}

\begin{table*}
    \small 
    \centering
    \begin{tabular}{lccc}
        \toprule
        \multirow{2}{*}{\textbf{Type}} & \multicolumn{3}{c}{\textbf{Example}} \\
        \cmidrule{2-4} 
        & \textbf{Input} & \textbf{Entity} & \textbf{Prediction} \\
        \midrule
        
                Noisy & Type of power plugs standardized by the National & \href{https://babelnet.org/synset?id=bn\%3A03357573n&orig=bn\%3A03357573n&lang=EN} {\textbf{bn:03357573n}} & \href{https://babelnet.org/synset?id=bn\%3A00030158n&orig=outlet&lang=EN} {\textbf{bn:00030158n}}\\
        $12\%$ &  Electrical Manufacturers Association & Nema Connector & Electric Outlet\\
        \specialrule{0em}{2pt}{2pt}
        & \includegraphics[width=0.10\linewidth]{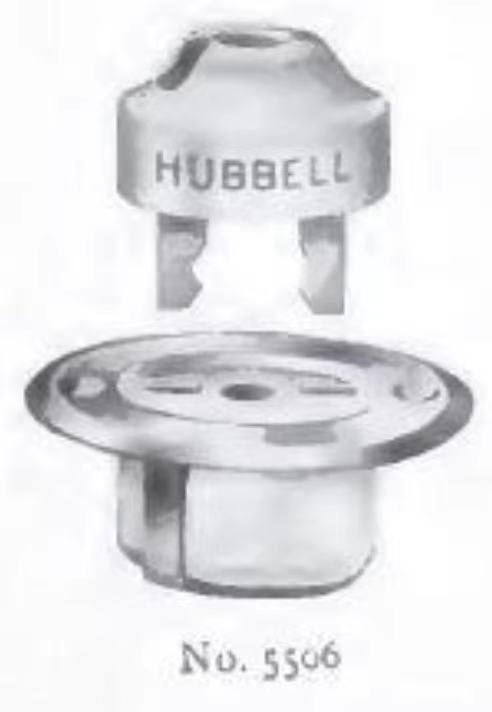} & & \\
        \midrule
        
        Hard & A town, and associated province &  \href{https://babelnet.org/synset?id=bn\%3A00665773n&orig=bn\%3A00665773n&lang=EN}{\textbf{bn:00665773n}} & 
        \href{https://babelnet.org/synset?id=bn\%3A00665687n&orig=bn\%3A00665687n&lang=EN}{\textbf{bn:00665687n}}\\
        $46\%$ & in Sardinia, Italy. & 
Province of Sassari & Province of Cagliari \\  
        \specialrule{0em}{2pt}{2pt}
        & \includegraphics[width=0.15\linewidth]{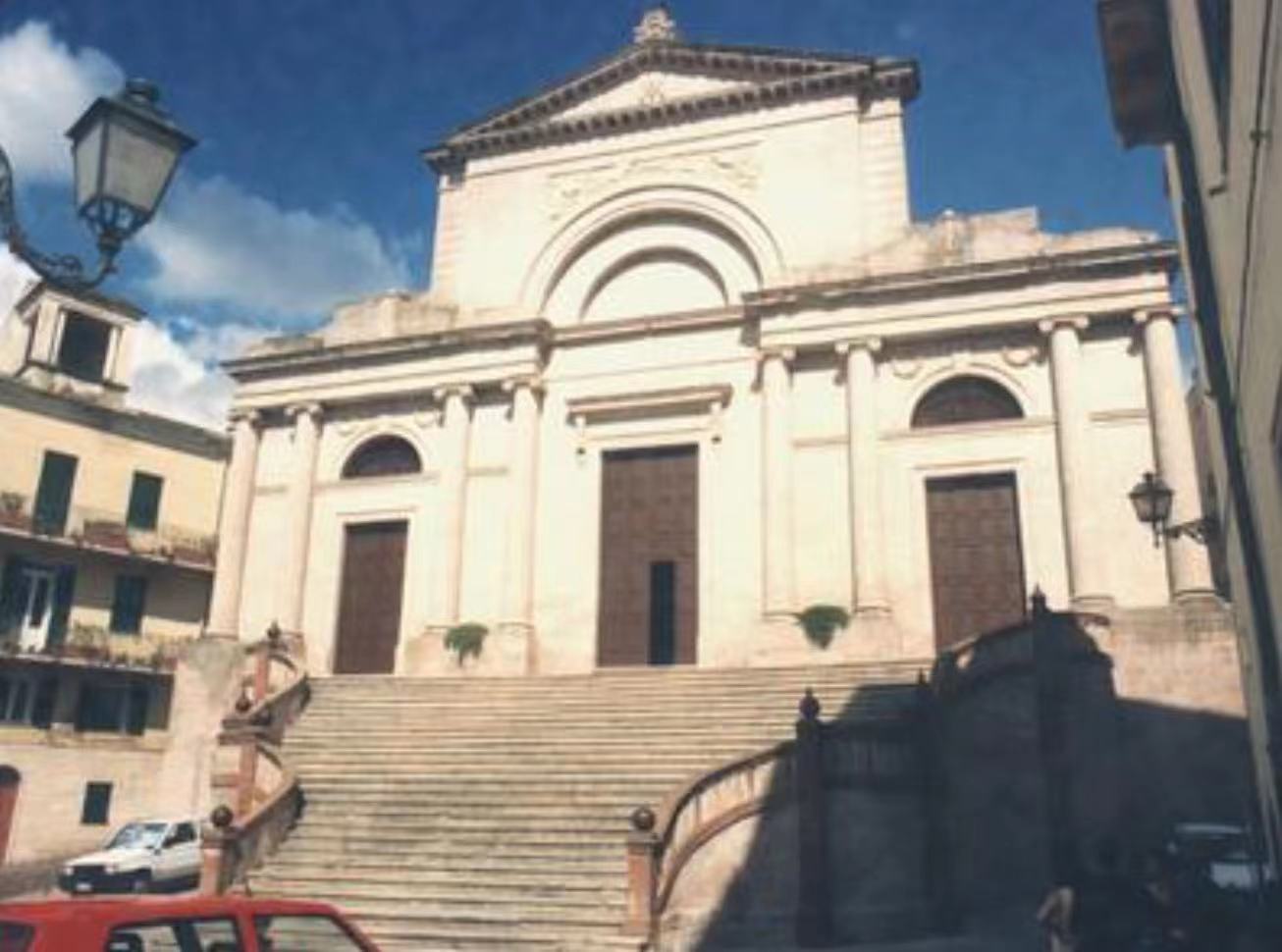} & & \\
        \midrule
        
        Wrong & A former aircraft maker, now PART & \href{https://babelnet.org/synset?id=bn\%3A01455566n&orig=bn\%3A01455566n&lang=EN}{\textbf{bn:01455566n}} & \href{https://babelnet.org/synset?id=bn\%3A01359190n&orig=bn\%3A01359190n&lang=EN}{\textbf{bn:01359190n}} \\ 
        $42\%$ & of Northrop Grumman  & Grumman & Ingalls Shipbuilding \\
        \specialrule{0em}{2pt}{2pt}
        & \includegraphics[width=0.15\linewidth]{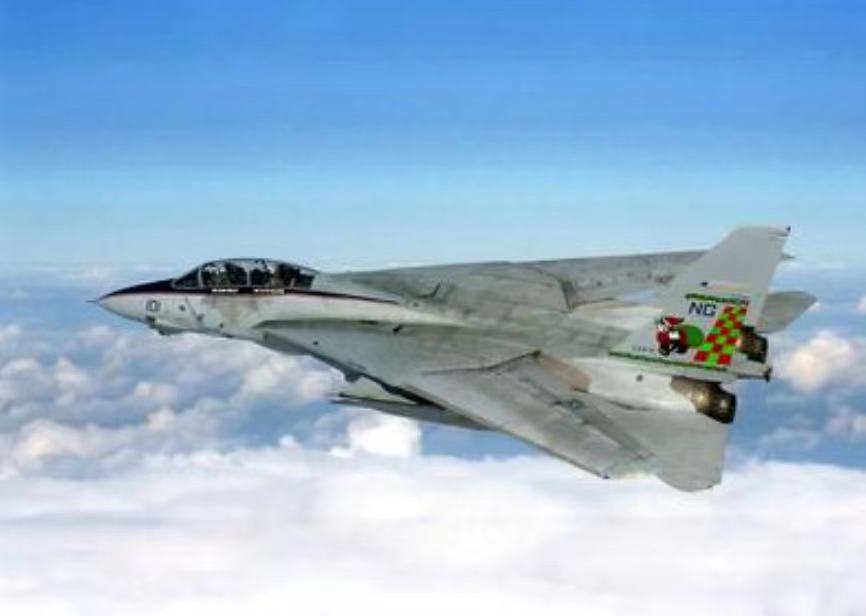} & & \\
        \bottomrule
    \end{tabular}
    \caption{Error analysis. 1) ``\textbf{Noisy}''. ``Electric Outlet'' has power plugs images associated which misleads models. 2) ``\textbf{Hard}''. The input image of ``Province of Sassari'' is ambiguous and hard to recognize the corresponding city. 3) ``\textbf{Wrong}''. The input text of ``Grumman'' is similar to ``Ingalls Shipbuilding'' and text matching models fail to recognize the correct entity ``Grumman'' which indicates more powerful models for multimodal representation are needed. 
    One can click entity ids to see details of the entities.}
    \label{tab:error_analysis}
\end{table*}


\begin{table*}
    \small
    \centering
    \begin{tabular}{ll}
    \toprule
    
    \href{https://babelnet.org/synset?id=bn\%3A00068195n&orig=bn\%3A00068195n&lang=EN}{\textbf{bn:00068195n}} &  \tabitem A style of architecture developed in Italy and western Europe between the Roman and \\
    Romanesque Architecture & the Gothic styles after 1000 AD;\\
    & \tabitem Characterized by round arches and vaults and by the substitution of piers for columns \\
    & and profuse ornament and arcades. \\ 
    & \tabitem Romanesque architecture is an architectural style of medieval Europe characterized \\ 
    & by semi-circular arches. \\
    & \tabitem Architecture of Europe which emerged in the late 10th century and lasted to the \\
    & 13th century. \\ 
    & \includegraphics[width=0.7\linewidth]{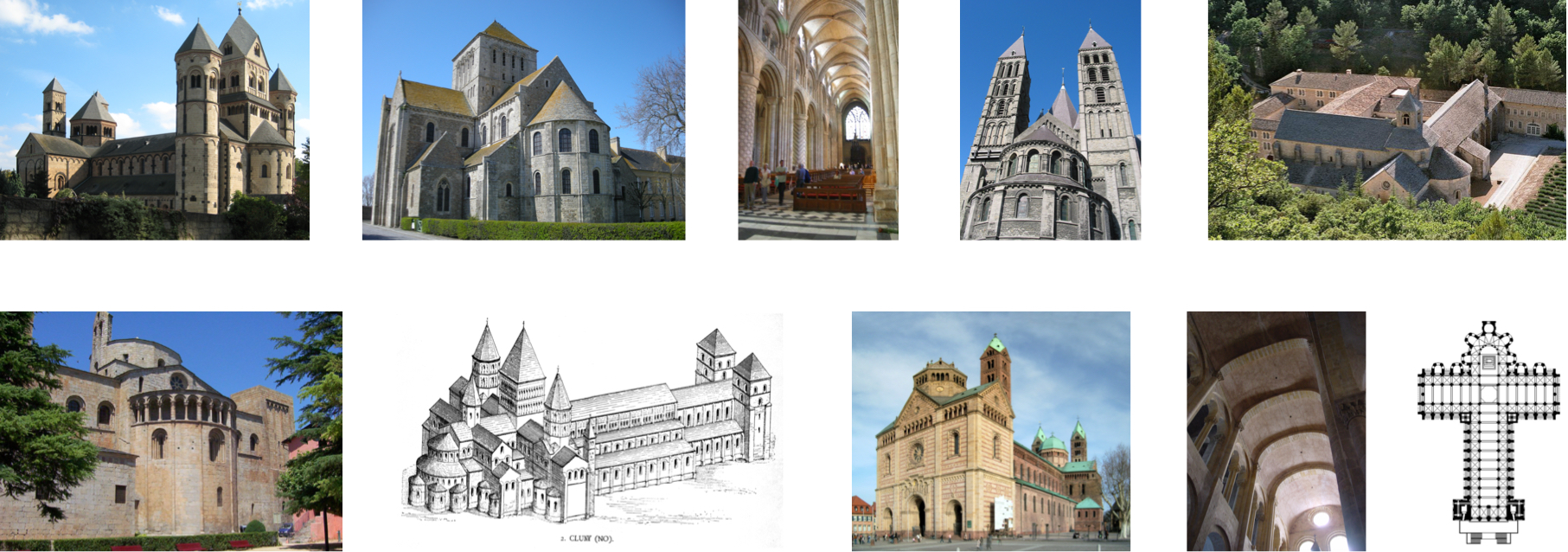} \\ 
    \midrule
    
    \href{https://babelnet.org/synset?id=bn\%3A00015540n&orig=bn\%3A00015540n&lang=EN}{\textbf{bn:00015540n}} & \tabitem The capital and largest city of France; and international center of culture and commerce. \\
    Paris & \tabitem Paris is the capital and most populous city of France, with an estimated population of \\ 
    & 2,148,271 residents as of 2020, in an area of more than 105 square kilometres. \\ 
    & \tabitem Designated nouvelle Rome at various stages of history between the reigns of Philip IV \\
    & and Louis XIV. \\ 
    & \tabitem The capital and largest city of France. \\
    & \tabitem Capital of France.  \\ 
    & \includegraphics[width=0.7\linewidth]{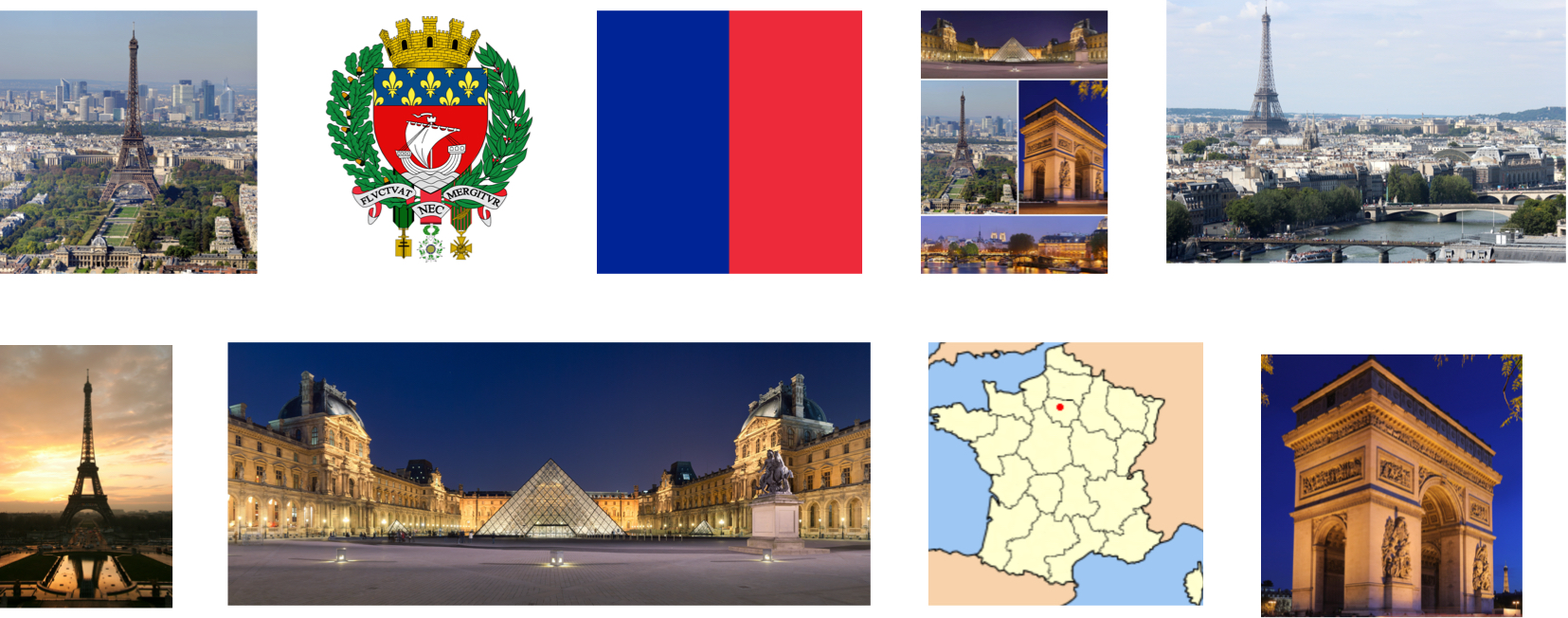} \\ 
    \midrule 

    \href{https://babelnet.org/synset?id=bn\%3A00036225n&orig=bn\%3A00036225n&lang=EN}{\textbf{bn:00036225n}} & \tabitem Francisco José de Goya y Lucientes was a Spanish romantic painter and printmaker. \\ 
    Francisco de Goya & \tabitem Spanish painter well known for his portraits and for his satires (1746-1828). \\ 
    & \tabitem 18th and 19th-century Spanish painter and printmaker. \\ 
    & \tabitem Francisco Goya, a Spanish painter. \\
    & \includegraphics[width=0.7\linewidth]{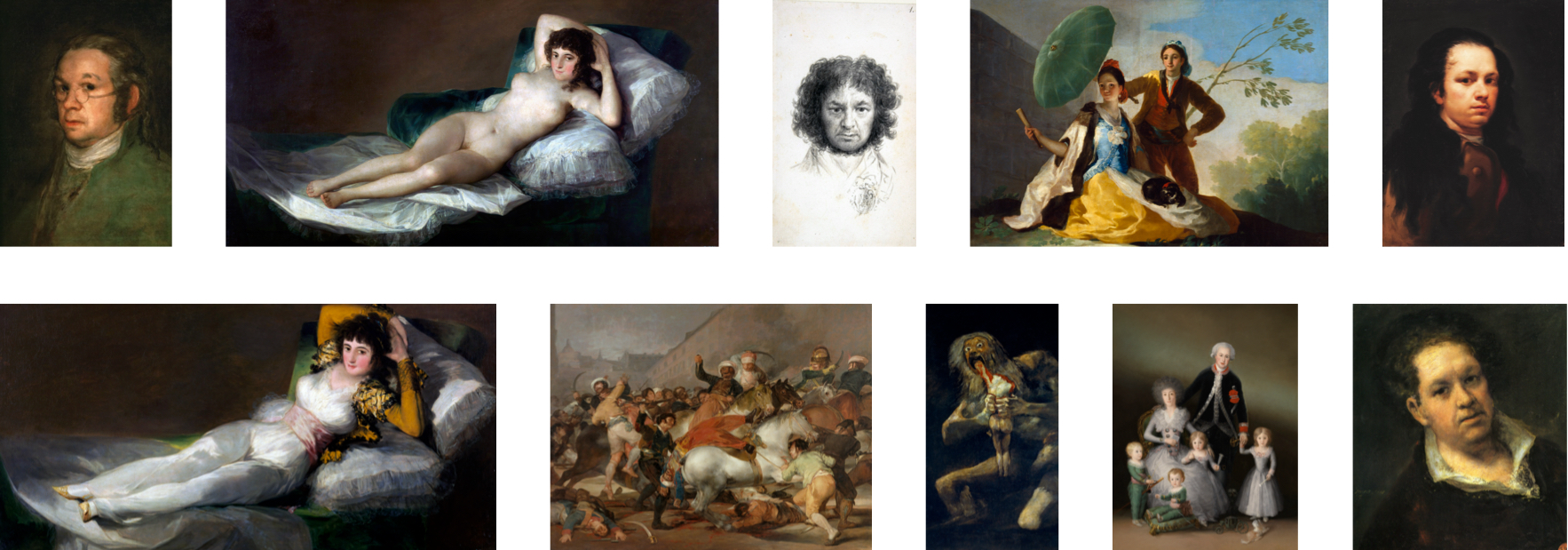} \\ 
    \bottomrule
    
    \end{tabular}
    \caption{Examples showing the diversity of data. The examples show typical entity categories: 1) \textit{Conceptual} entities containing various instances and images. 2) \textit{Location} entities containing diverse images related to the location (e.g., landmarks, maps, symbols). 3) \textit{Person} entities containing diverse images related to the person (e.g., portraits, works). }
    \label{tab:diversity}
\end{table*}

\begin{table*}
    \small
    \centering
    \begin{tabular}{cll}
    \toprule
    \textbf{Image} & \textbf{Text} & \textbf{Entity} \\ 
    
    \midrule
    \multirow{4}{*}{\includegraphics[width=0.09\linewidth]{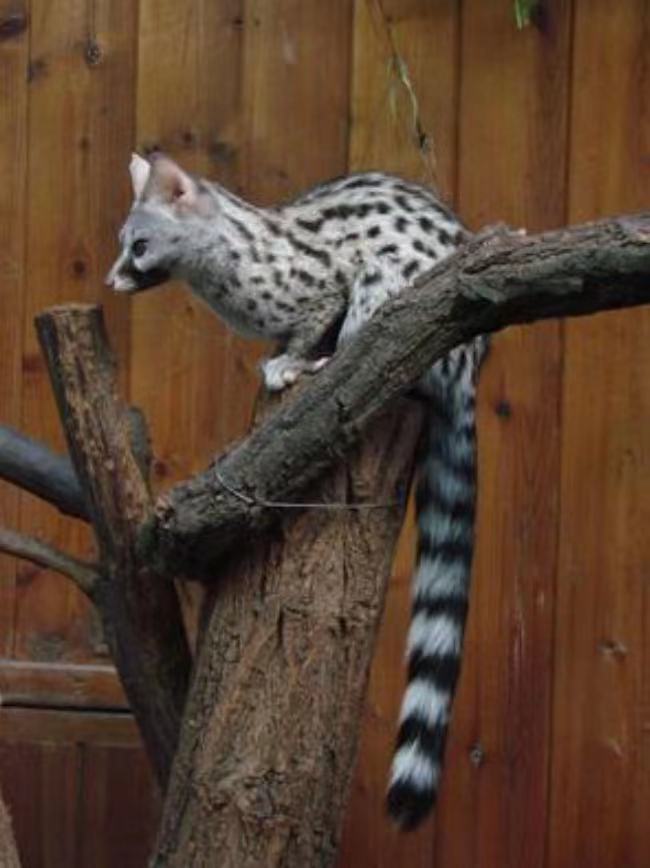}} & The type genus of the family Procyonidae: raccoons. & \href{https://babelnet.org/synset?id=bn\%3A00039337n&orig=bn\%3A00039337n&lang=EN}{\textbf{bn:00039337n}} (Procyon) \\ 
    \specialrule{0em}{2pt}{2pt}
    & Small cat-like predatory mammals of warmer parts of the Old World. & \href{https://babelnet.org/synset?id=bn\%3A00080161n&orig=bn\%3A00080161n&lang=EN}{\textbf{bn:00080161n}} (Viverrine) \\ 
    \specialrule{0em}{2pt}{2pt}
    & Any animal of the phylum Chordata having a notochord or spinal column.  & \href{https://babelnet.org/synset?id=bn\%3A00018748n&orig=bn\%3A00018748n&lang=EN}{\textbf{bn:00018748n}} (Chordate)\\ 
    \specialrule{0em}{2pt}{2pt}
    & Agile Old World viverrine having a spotted coat and long ringed tail. & \href{https://babelnet.org/synset?id=bn\%3A00037694n&orig=bn\%3A00037694n&lang=EN}{\textbf{bn:00037694n}} (Genet) \\ 
    \specialrule{0em}{2pt}{2pt}

    \cmidrule{2-3} 
    \multirow{2}{*}{\includegraphics[width=0.1\linewidth]{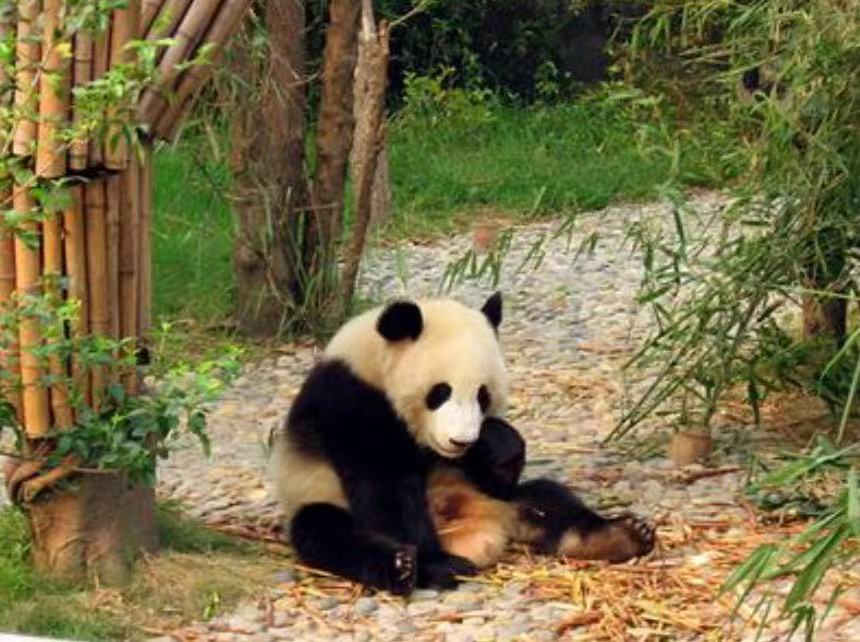}} & Large black-and-white herbivorous mammal of bamboo forests of China. & \href{https://babelnet.org/synset?id=bn\%3A00002174n&orig=bn\%3A00002174n&lang=EN}{\textbf{bn:00002174n
    }} (Giant Panda) \\ 
    \specialrule{0em}{2pt}{2pt}
    & Woody tropical grass having hollow woody stems; mature canes used & \\ 
    & for construction and furniture. & \href{https://babelnet.org/synset?id=bn\%3A00008254n&orig=bn\%3A00008254n&lang=EN}{\textbf{bn:00008254n}} (Bamboo) \\ 
    \specialrule{0em}{2pt}{2pt}

    \cmidrule{2-3} 
    \multirow{8}{*}{\includegraphics[width=0.1\linewidth]{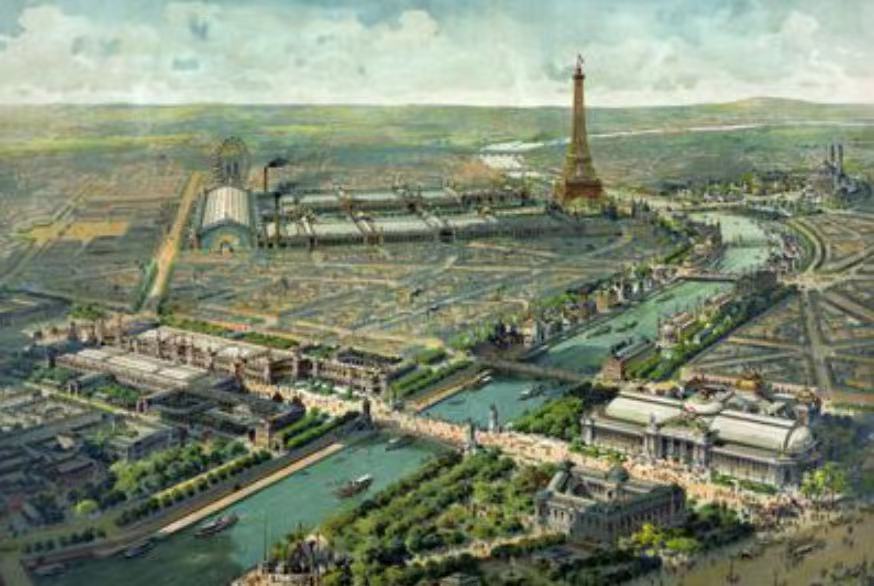}} & A situation or topic as if viewed from an altitude or distance. & \href{https://babelnet.org/synset?id=bn\%3A00010616n&orig=bn\%3A00010616n&lang=EN}{\textbf{bn:00010616n}} (Bird's Eye View) \\
    \specialrule{0em}{2pt}{2pt}
    & A wrought iron tower 300 meters high that was constructed in Paris & \\
    & in 1889; for many years it was the tallest man-made structure. & \href{https://babelnet.org/synset?id=bn\%3A00029980n&orig=bn\%3A00029980n&lang=EN}{\textbf{bn:00029980n}} (Eiffel Tower) \\
    \specialrule{0em}{2pt}{2pt}
    & The capital and largest city of France; and international center & \\ 
    & of culture and commerce. & \href{https://babelnet.org/synset?id=bn\%3A00015540n&orig=bn\%3A00015540n&lang=EN}{\textbf{bn:00015540n}} (Paris) \\ 
    \specialrule{0em}{2pt}{2pt}
    & An art movement launched in 1905 whose work was characterized & \\
    & by bright and nonnatural colors and simple forms; influenced the & \\
    & expressionists & \href{https://babelnet.org/synset?id=bn\%3A00033829n&orig=bn\%3A00033829n&lang=EN}{\textbf{bn:00033829n}} (Fauvism) \\
    \specialrule{0em}{2pt}{2pt}
    
    \bottomrule
    \end{tabular}
    \caption{Examples showing the ambiguity in the data. One image may contain 1) various entities. The second image contains entities ``Giant Panda'', ``Bamboo''. The third image contains entities ``Eiffel Tower'', ``Paris'', ``Fauvism''. If not specified, it is hard to recognize the corresponding entity. 2) an object corresponding to various entities. The first image contains an animal object. If considering different species categories level, the entity can be ``Viverrine'', ``Chordate'', ``Genet''. 
    3) rare entities. The third image contains entity ``Bird's Eye View'' which is obscure and rare.}
    \label{tab:ambiguity}
\end{table*}



\end{document}